\documentclass[11pt, a4paper]{article}
\usepackage{jcappub}

\title{Dynamical templates for comparison to $\Lambda$CDM:
Static or dynamical dark energy?}
\author{Rodger I. Thompson}
\affiliation{Steward Observatory and Department of Astronomy, University of Arizona\\
933 N. Cherry Ave., Tucson Arizona, USA}
\emailAdd{rit@email.arizona.edu}

\abstract{This study forges new tools to discriminate between dynamical and static dark energy.  It provides accurate evolutionary 
templates of dynamical cosmological parameters and fundamental constants as analytic functions of the scale factor.  
They are designed to replace the commonly used parameterizations in likelihood calculations with evolutionary templates 
based on the physics of specific dynamical cosmologies and dark energy potentials. Thus they are termed Cosmology and
Potential Specific, CPS, templates. A suite of CPS templates are calculated for a flat quintessence cosmology with a dark
energy potential of the same mathematical form as the Higgs potential.  This Higgs inspired, HI, polynomial potential produces
a rich set of evolutions unlike most monomial  potentials.  The study produces CPS templates that are analytic functions of
the scale factor. It uses a recently developed beta function formalism that provides a differential function for the scalar in
terms of the scale factor. This establishes a methodology for easily producing CPS templates for other dark energy potentials
and cosmologies to determine the likelihoods of dynamical cosmologies relative to $\Lambda$CDM. The study also 
examines the evolution of fundamental constants such as the proton to electron mass ratio and the fine structure constant  
involving an intersection between particle physics and cosmology. Appendix~\ref{as-A} displays an abridged suite of 
CPS templates for flat quintessence and the HI dark energy potential.}


\begin{document}

\maketitle

\section{Introduction}\label{s-intro}
The decadal review of astrophysics and astronomy for the 2020s \cite{nata21} identifies the question of whether the dark
energy density of the universe is static or dynamic as one of the great unanswered questions in modern science. The answer 
requires an accurate comparison of the relative likelihoods of $\Lambda$CDM and dynamical cosmologies.  The predicted 
evolutions of cosmological parameters, templates, for $\Lambda$CDM  are well known based on general relativity and the 
standard model of physics.  Calculations of the likelihood of dynamical cosmologies, however, are most often based on 
parameterizations rather than the physics of the cosmologies and dark energy potentials.  Proper scientific inquiry requires
likelihoods based on the physics of the cosmology and dark energy potential that is being tested.  This study presents
a methodology for achieving that goal. It demonstrates the methodology by calculating evolutionary templates for a 
specific dynamical cosmology and dark energy potential that provide significant improvements over the current 
parameterizations.  The need for Cosmology and Potential Specific, CPS, templates is  discussed in section~\ref{s-ncps} in 
terms of the current parameterization for the dark energy equation of state, EoS. Evolutions of cosmological parameters such 
as the dark energy EoS, $w$, are strong functions of both the cosmology and the dark energy potential. The multiplicity of 
dynamical cosmologies and number of possible dark energy potentials has been one of the sticking points hindering the 
development of dynamical templates. The methodology in this work simplifies calculating accurate evolutionary CPS 
templates.  The well studied flat quintessence dynamical cosmology provides a logical starting point for the study.  The
dark energy potential has the same mathematical form as the Higgs potential, $V(\phi) \propto (\phi^2-\gamma^2)^2$, and 
is therefore referred to as the Higgs Inspired or HI potential in subsequent discussion.  The value of the constant $\gamma$ 
is greater than the current scalar $\phi_0$ placing the equilibrium point $\phi = \gamma$ in the future.  At early times
the constant $\gamma^4$ term dominates, acting as a cosmological constant until the dynamical terms produce significant
evolution during the late times, $z \le 10$, considered here. The detailed properties of the HI potential are examined in 
section~\ref{ss-hipot}. 

For likelihood calculations the CPS templates should be accurate, analytic and functions of an observable cosmological 
variable such as the scale factor $a$ or redshift $z$.  The scale factor $a = 0.1-1.0$ is the  variable used in this study. The 
recently developed beta function formalism \cite{bin15,cic17,koh17}, presented in section~\ref{s-bff}, is the key to templates 
that are  functions of the scale factor $a$.  It provides a methodology for creating the templates in a simple manner that 
overcomes the problem of the multitude of dynamical cosmologies and dark energy potentials. Extension of the methodology 
to DBI and non-standard kinetic term cosmologies has been accomplished by \cite{bin17}.  Cosmologies with non-minimal 
couplings have been also addressed \cite{cic17}.  The formalism has its origins in particle physics renormalization studies 
\cite{cic17}. The cosmological beta function is a function of the scalar $\phi$ and the scale factor $a$ that provides a 
differential equation to solve for $\phi(a)$. The details of the beta function formalism in a quintessence cosmology are 
presented in section~\ref{s-bff}.

The primary purpose of CPS templates is likelihood calculations.  They, however, play important
roles in other areas such as observation planning where accurate templates indicate the best redshift regions for
measurements and the accuracy needed to discriminate between static and dynamical dark energy. They also indicate which 
cosmological parameters are sensitive measures of dark energy and which are not. The range of evolution is also indicated 
by the CPS templates such as the perhaps unexpected non-monotonic evolution of the dark energy EoS for some of the cases 
in this study.  CPS templates also serve as the initial solution for numerical exact solutions to the cosmology and potential.
This may occur for cosmologies and potentials that have a high likelihood.

Natural units are used throughout with $c$, $\hbar$ and $8\pi G$ set to one.  Masses are given in terms of reduced
Planck masses, $M_p$ and the constant $\kappa$ is defined as $\frac{1}{M_p}$. In the adopted mass units $\kappa$
is equal to one but is retained in equations and relevant text to properly indicate the units.  

\section{The specific cosmology and potential}\label{s-scp}
The specific cosmology and potential in this study is  flat quintessence with a dark energy potential that has the mathematical 
form of the Higgs potential.  Quintessence is well developed \cite{cop06, bah18} and a basic example of a dynamical 
scalar field dark energy cosmology. The HI potential's mathematical relation to the Higgs has a degree of naturalness that
makes it .physically motivated. The analysis and CPS templates developed are based on the physics of this specific
cosmology and potential rather than a parameterization.

\subsection{Flat quintessence}\label{ss-q}
A few quintessence features used in this study are given here for convenience. The quintessence dark energy density, 
 $\rho_{\phi}$, and pressure, $P_{\phi}$ equations are in terms of the scalar $\phi$ with units of mass and the potential 
 with units of mass$^4$.
\begin{equation} \label{eq-rhop}
\rho_{\phi} \equiv \frac{\dot{\phi}^2}{2}+V(\phi), \hspace{1cm} P_{\phi} 
\equiv \frac{\dot{\phi}^2}{2}-V(\phi)
\end{equation}
where $\dot{\phi}$ has units of mass squared.  The dark energy EoS $w(\phi)$ is
\begin{equation} \label{eq-deos}
w(\phi)= \frac{P_{\phi}}{\rho_{\phi}}=\frac{\frac{\dot{\phi}^2}{2}-V(\phi)}{\frac{\dot{\phi}^2}{2}+V(\phi)}.
\end{equation}
Combining eqns.~\ref{eq-rhop} gives 
\begin{equation}\label{eq-pdots}
P_{\phi} + \rho_{\phi}  =\dot{\phi}^2
\end{equation}
It follows that
\begin{equation}\label{eq-wpo}
\frac{P_{\phi} +\rho_{\phi}}{\rho_{\phi}} = w+1=\frac{\dot{\phi}^2}{\rho_{\phi}}
\end{equation}
giving $\dot{\phi}$ a relationship to the dark energy EoS and the dark energy density.
\begin{equation} \label{eq-dpw}
\dot{\phi}^2=\rho_{\phi}(w+1)
\end{equation}

\subsubsection{The Friedmann constraints}\label{sss-frcon}
In addition to the quintessence specific relations the two Friedmann constraints play an important role in the methodology.
The forms of the first and second Friedman constraints used here are
\begin{equation}\label{eq-friedcs}
3H^2(a) = \rho_{\phi}(a) +\rho_m(a)  \hspace{1cm} 3(\dot{H}(a) + H(a)^2)=-\frac{\rho(a)+3P(a)}{2}
\end{equation}
where $\rho_{\phi}(a)$ is the dark energy density, $\rho_m(a)$ is the matter density, $\rho(a)$ is the sum of the matter and 
dark energy densities and $P(a)$ is the dark energy pressure.

\subsection{The HI potential}\label{ss-hipot}
The model HI potential is given in a modified Ratra Peebles format \cite{rat88, pee88}. 
\begin{equation}\label{eq-hip}
V(\theta) = M^4 ((\kappa\theta)^2 - (\kappa\delta)^2)^2 = M^4((\kappa\theta)^4 -2(\kappa\theta)^2(\kappa\delta)^2 
+ (\kappa\delta)^4)
\end{equation} 
The constant $M$ has units of mass and contains all of the dimensionality of the potential. The true scalar $\phi$ is related to
$\theta$ by $\phi = M\kappa\theta$ and the true constant $\gamma$ is related to $\delta$ by $\gamma = M\kappa\delta$. 
The designations $\theta$ and $\delta$ are used to indicate that they are not the true scalar and constant.  $M$ is not a free 
parameter but is determined by the boundary conditions and the first Friedmann constraint. The scalar $\theta$ and the 
constant $\delta$ also have units of mass, hence $\kappa\theta$ and $\kappa\delta$ are dimensionless, eliminating the 
need for the $n$ in the usual $ M^{4+n}$ notation of the Ratra-Peebles format.  In this format $\kappa\theta$ and $\kappa\delta$ 
have magnitudes on the order of unity while $M$ is on the order of $10^{-31} M_p$. The format was partially chosen to make
 the derived CPS templates compatible with multiple standard mathematical platforms although the Mathematica platform used 
 in this study easily handles the order of $10^{-121}$ values encountered in the calculations. The true scalar $\phi$ 
in the potential is the dark energy scalar field with none of the complex physics associated with the Higgs field. The remainder 
of the study utilizes $\theta$ and $\delta$ to be consistent with the CPS templates.  

Since $\delta >\theta$ the constant $(\kappa\delta)^4$ term is dominant.  The scalar $\theta$ is much smaller than $\delta$ at 
early times therefore the $(\kappa\delta)^4$ term acts like a cosmological constant in early evolution.  The dynamical
 $-2(\kappa\theta)^2(\kappa\delta)^2$ term is the next largest term and provides the rolling of the potential to smaller values at 
 later times.  For the scale factor range considered here the dynamical $(\kappa\theta)^4$ is always subdominant but can 
 influence the second derivative of the evolution.  

\section{The Need for CPS Templates} \label{s-ncps}
An important example of the need for CPS templates is the ubiquitous Chevallier, Polarski, Linder, CPL, \cite{che01,lin03} 
linear parameterization of $w$ given by
\begin{equation}\label{eq-cpl}
w(a) = w_o+(1-a)w_a
\end{equation} 
CPL is the preferred parameterization of $w$   for a majority of new missions and facilities e.g. \cite{mar21}. 

\begin{figure}
\scalebox{.9}{\includegraphics{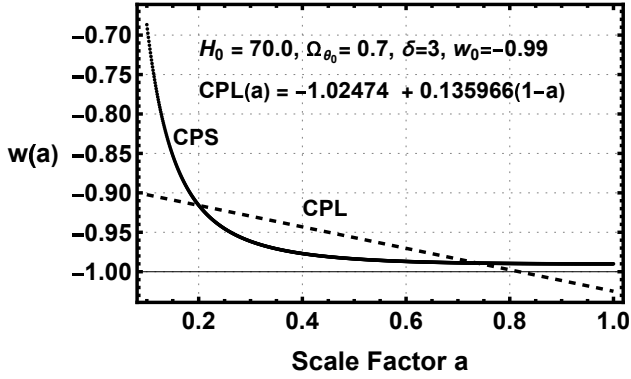}}
\caption{The dashed line is the CPL linear fit to the $w(a)$ freezing evolution for $\delta=3$ and $w_0=-0.99$, solid line,
from this work.}
\label{fig-cpl}
\end{figure}

Its linear form is quite adequate for testing the validity of $\Lambda$CDM and any deviation of $w_0$ from minus one
or $w_a$ from zero indicates a possible deviation from $\Lambda$CDM. Utilization of CPL for highly non-linear 
evolutions can, however, lead to erroneous conclusions.  To demonstrate this the CPL parameterization was fit
to a noiseless $w$ evolution for the $\delta = 3$, $w_0=-0.99$ case from this work.  Figure~\ref{fig-cpl} shows the results
with the calculated $w$ evolution as the solid line and the CPL fit as the dashed line.
The CPL fit correctly indicates that $\Lambda$CDM is not a good fit to the evolution, but it does not contribute information
about the actual evolution except that it is becoming more negative with time. More importantly the fit gives phantom values, 
$w < -1$, at scale factors greater than 0.8 for an evolution that has no phantom values. 

Three recent MCMC analyses of observational data \cite{pla20, div20, div21} with a CPL template find phantom values of
$w$ at low redshift. These results are interpreted by \cite{div20} as inconsistent with quintessence cosmologies since it is  
difficult for a quintessence cosmology to cross the phantom boundary \cite{vik05}. This example highlights the need for 
accurate, analytic CPS templates rather than parameterizations of cosmological parameter evolution in dynamical cosmologies.
In this case a possibly inaccurate conclusion is found based on a parameterization misfit rather than on physics based 
predictions. Analyses of the current and  the expected new more precise and sensitive cosmological observations need 
accurate templates for proper interpretation of the data. This work concentrates on providing them.

\section{The beta function formalism}\label{s-bff}
The beta function formalism is based on  commonalities between cosmological
inflation and the particle physics based Renormalization Group, RG, flow  \cite{bin15,cic17,koh17}.  In the formalism the beta function is a relationship 
between the dynamical scalar $\theta$ that is the source of dark energy and the scale factor $a$.  In \cite{bin15,cic17,koh17}
beta is set to a common function such as a power law or an exponential and the resulting evolutions are explored.  Here
the potential is set and an appropriate beta function based on the potential is calculated to provide accurate evolutions for a 
specific cosmology and potential. The key is to use the beta function to determine $\kappa\theta(a)$ so that evolutions
expressed in terms of the scalar are expressed in terms of the scale factor $a$.  As described in the following this requires
replacing the model potential with a potential, beta potential, that accurately but not exactly represents the model potential.

\subsection{The beta function}\label{ss-bf}
The beta function in the flat quintessence cosmology formalism is defined as
\begin{equation}\label{eq-beta}
\beta(\kappa\theta) \equiv \frac{ d(\kappa\theta)}{d\ln(a)} = a\frac{d(\kappa\theta)}{da}
\end{equation}
which defines a differential equation for $\theta(a)$ dependent on the form of $\beta(\theta)$.  Even in this simple form
it is apparent that
\begin{equation}\label{eq-bdat}
\frac{d(\kappa\theta)}{da} = \frac{\beta(\kappa\theta}{a}, \hspace{1cm} \frac{d(\kappa\theta)}{dt} = H \beta =
\frac{1}{a}\frac{d a}{dt}a\frac{d(\kappa\theta)}{da} =\kappa\dot{\theta}
\end{equation}
from the definition of the Hubble parameter $H$. The first Friedmann constraint, see section~\ref{sss-frcon}, and 
equation~\ref{eq-bdat} give
\begin{equation}\label{eq-bvfc}
3 H^2 =\rho_{\theta} = \frac{\dot{\theta}^2}{2} +V(\theta) = \frac{\beta^2 H^2}{2} +V(\theta)
\end{equation}
where $\rho_{\theta}$ is the dark energy density.  In the presence of mass $3H^2$ is replaced by $3\Omega_{\theta}H^2$.  
The potential $V(\theta)$ is then
\begin{equation}\label{eq-bv}
V(\theta) =3H^2- \frac{\beta^2 H^2}{2} = 3H^2(1-\frac{\beta^2}{6}) = \rho_{\theta}(1-\frac{\beta^2}{6})
\end{equation}
From \cite{cic17} the potential and beta function are related by
\begin{equation}\label{eq-bvbc}
V(\theta)=\exp\{-\int \beta(\theta')d\theta'\}(1-\frac{\beta^2}{6})
\end{equation}
that requires that either beta be the negative of the logarithmic derivative of the dark energy density or that the
potential is defined by the beta function in equation~\ref{eq-bvbc}. In general and in this specific case the integral
equation that defines beta as the negative of the logarithmic derivative of the dark energy density does not have
an analytic solution.  Equation~\ref{eq-bvbc},however provides the opportunity
to meet the goal of accurate analytic evolutionary templates.  Setting the beta function to the negative of the 
logarithmic derivative of the model potential provides solutions for a cosmology with a dark energy potential that
is the model potential multiplied by $(1-\beta^2/6)$.  If $\beta^2/6$ is small relative to one accurate but not exact
solutions that satisfy the criterion of significant improvement over parameterizations are achieved.  The evolutions are 
then exact for the beta potential and cosmology and are accurate for the model potential at an accuracy of  
$(1-\frac{\beta^2}{6})$.  An accuracy threshold of $99\%$ is set for this study. For all of the cases in the study the 
beta potential matches the model potential to an accuracy of $99.6\%$ or better.

\subsubsection{The HI potential beta function}\label{sss-hibeta}
The model potential for the beta function is equation~\ref{eq-hip}.  Taking the negative of the logarithmic derivative of
the model potential produces the beta function
\begin{equation}\label{eq-bvde}
\beta(\kappa \theta)=\frac{d\kappa\theta}{d\ln(a)}=\frac{-4\kappa\theta}{(\kappa\theta)^2-(\kappa\delta)^2}
\end{equation}
giving a differential equation for $\theta(a)$.
$\beta(\kappa\theta)$ is not a function of $M$ leaving $M$ free to satisfy the boundary conditions.
Equation~\ref{eq-bvde} shows that the beta function is positive since $\delta > \theta$, therefore, $\theta$ is increasing 
toward $\delta$.  The beta function links the evolution of the scalar to the evolution of the scale factor via the differential 
equation~\ref{eq-bvde}.  Using $\dot{\theta} =\beta H$ and $\rho_{\theta} = 3\Omega_{\theta}H^2$ equations~\ref{eq-dpw} 
and~\ref{eq-bvde} show that
\begin{equation}\label{eq-bw}
\beta (\kappa\theta)=\sqrt{3\Omega_{\theta}(w+1)}
\end{equation}
Using the current values of the variables and parameters in equations~\ref{eq-bvde} and~\ref{eq-bw} sets the current 
value of $\kappa\theta$.
\begin{equation}\label{eq-to}
\kappa\theta_0=-\frac{4-\sqrt{16+12\Omega_{\theta_0}(w_0+1)(\kappa\delta)^2}}{2\sqrt{3\Omega_{\theta_0}(w_0+1)}}
\end{equation}
the positive solution of the equation since the square root term in the numerator is greater than four. 

\section{Parameter evolution derivations}\label{s-ped}
The beta function~\ref{eq-bvde} provides the means to derive the cosmological parameters as a function of the scale factor. 
Both Friedman constraints, section~\ref{sss-frcon}, are utilized in the derivations of the cosmological parameters.   The first 
step is the derivation of $\kappa\theta(a)$. Section~\ref{ss-tofa} shows that $\theta(a)$ is a function of the Lambert W function 
$W(a)$.  Parameters of $\theta(a)$ are also derived as functions of $W(a)$ by substituting the proper function of $W(a)$ for 
$\theta(a)$.  The resulting functions of $W(a)$ are reduced from the pure substitution to simplified expressions, if possible, 
without explicitly displaying the intermediate calculations to save space.

\subsection{$\kappa\theta$ as a function of the scale factor}\label{ss-tofa}
The analysis in this section utilizes the quintessence relationships described in section~\ref{ss-q} and is therefore strictly
valid only for flat quintessence. Rearranging equation~\ref{eq-bvde} gives 
\begin{equation}\label{eq-yb}
4 d(\ln(a))=-\frac{(\kappa\theta)^2-(\kappa\delta)^2}{\kappa\theta}d(\kappa\theta)
\end{equation}
Integrating both sides of eqn.~\ref{eq-yb} over 1 to $a$ on the left and $\theta_0$ to $\theta$ on the right gives
\begin{equation}\label{eq-intb}
8\ln(a)=2 (\kappa\delta)^2\ln(\kappa\theta)-(\kappa\theta)^2-(2 (\kappa\delta)^2\ln(\kappa\theta_0)-\kappa\theta_0^2)
\end{equation}
where $\kappa\theta_0$ is given in equation~\ref{eq-to}.

Setting the constant  $2 (\kappa\delta)^2\ln(\kappa\theta_0)-\kappa\theta_0^2$ to $c$ and dividing both sides by 
$(\kappa\delta)^2$ yields
\begin {equation}\label{eq-bg}
\frac{8}{(\kappa\delta)^2}\ln(a)+\frac{c}{(\kappa\delta)^2}=2\ln(\kappa\theta)-\frac{(\kappa\theta)^2}{(\kappa\delta)^2}
\end{equation}

Taking the exponential of both sides of eqn.~\ref{eq-bg} and dividing by $-(\kappa\delta)^2$  produces
\begin{equation}\label{eq-ya}
-\frac{a^{\frac{8}{(\kappa\delta)^2}}}{(\kappa\delta)^2}e^{\frac{c}{(\kappa\delta)^2}}=-(\frac{\theta}{\delta})^2e^{-(\frac{\theta}{\delta})^2}
\end{equation}
Equation~\ref{eq-ya} has the form of  the Lambert W function which is  the solution to the equation
\begin{equation}\label{eq-W}
x= W(x)e^{W(x)}
\end{equation}
where here
\begin{equation}\label{eq-wx}
x=-\frac{a^{\frac{8}{(\kappa\delta)^2}}}{(\kappa\delta)^2}e^{\frac{c}{(\kappa\delta)^2}} \hspace{1cm} W(x) = -(\frac{\theta}{\delta})^2
\end{equation}
The Lambert W function commonly appears in cosmology and general relativity so its appearance here has a degree of
naturalness.

The analytic solution for $\kappa\theta(a)$ is
\begin{equation}\label{eq-ywa}
\kappa\theta(a)=\kappa\delta\sqrt{-W(-\frac{a^{\frac{8}{(\kappa\delta)^2}}}{(\kappa\delta)^2}e^{\frac{c}{(\kappa\delta)^2}})}
\end{equation}
that is the positive solution for the square root which is real since $W(x)$ is negative. 

Negative values of $W(x)$ occur if $x$ is between $-\frac{1}{e}$ and zero \citep{olv10}, satisfied for all cases in this work. 
A concise form of equation~\ref{eq-ywa} is
\begin{equation}\label{eq-wqap}
\kappa\theta(a)=\kappa\delta\sqrt{-W(q a^{\frac{8}{(\kappa\delta)^2}})}
\end{equation}
where $a$ is the scale factor and the constant $q$ is
\begin{equation}\label{eq-qp}
q=-\frac{e^{\frac{c}{(\kappa\delta)^2}}}{(\kappa\delta)^2}
\end{equation}
An even more concise notation is achieved by defining
\begin{equation}\label{eq-wchi}
\chi = q a^{\frac{8}{(\kappa\delta)^2}}
\end{equation}
The scalar is then
\begin{equation}\label{eq-thchi}
\kappa\theta = \kappa\delta \sqrt{-W(\chi)}
\end{equation}
\begin{figure}
\scalebox{.9}{\includegraphics{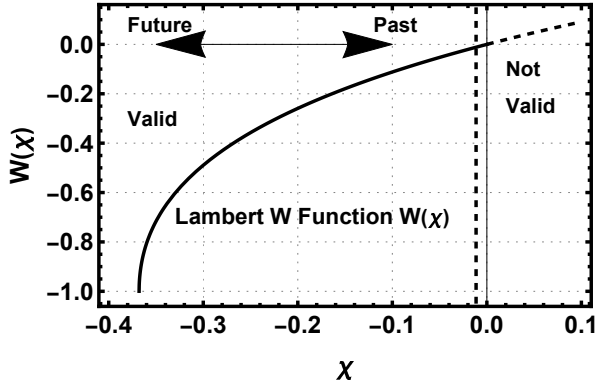}}
\caption{The Lambert W function showing the evolution for negative arguments which produce valid solutions
for $\theta$ and positive arguments that produce invalid solutions, dashed line. The region between zero and
the vertical dashed line at -0.0114 is the total range of $\chi$ for all of the cases in this study. The double arrow
indicates the directions of future and past evolutions.}
\label{fig-lamw}
\end{figure}
Figure~\ref{fig-lamw} shows the valid and invalid regions of the Lambert W function in eqn.~\ref{eq-ywa}.  The value
of $\chi$ is moving to the left with time in the figure after starting just negative of zero as indicated by the double arrow
in the figure. The vertical dashed line shows the maximum excursion of $\chi$ for the cases in this study. Only a small portion 
of the valid $\chi$ values are occupied by the past leaving a large range for future evolution.

\subsubsection{The beta function in terms of the Lambert W function}\label{sss-betaw}
The beta function as a function of $\theta(a)$ is equation~\ref{eq-bvde}.  Expressing it as a function of $W(\chi)$
demonstrates some of the common functional elements in terms of $W(\chi)$.  The Lambert W function expression
for $(\kappa\theta)^2$ is
\begin{equation}\label{eq-th2w}
(\kappa\theta)^2  = -(\kappa\delta)^2W(\chi)
\end{equation}
The often appearing $(\kappa\theta)^2 - (\kappa\delta)^2$ is then
\begin{equation}\label{eq-kdw}
(\kappa\theta)^2 - (\kappa\delta)^2 = -(\kappa\delta)^2W(\chi) -(\kappa\delta)^2 = -(\kappa\delta)^2(W(\chi)+1)
\end{equation}
resulting in a beta function of
\begin{equation}\label{eq-betaw}
\beta(W(\chi)) =\frac{4\sqrt{-W(\chi)}}{\kappa\delta(W(\chi)+1)}
\end{equation}
The scalar form of $(1-\beta^2/6)$ is
\begin{equation}\label{eq-bcorth}
1-\frac{\beta^2(\theta)}{6} = 1-\frac{8(\kappa\theta)^2}{3((\kappa\theta)^2-(\kappa\delta)^2)^2}
\end{equation}
The Lambert W function form for $(1-\beta^2/6)$ is
\begin{equation}\label{eq-bcorw}
1-\frac{8W(\chi)}{3\delta^2(W(\chi)+1)^2}
\end{equation}

\subsubsection{Calculation of the potential constant $M$}\label{sss-calm}
At this point two potentials are defined, the model HI potential defined by equation~\ref{eq-hip} and the beta potential
that is $(1-\frac{\beta^2}{6})$ times the model potential.  Their $M$ parameters are termed $M_{mod}$ and $M_{\beta}$
respectively.  In general the CPS templates are created using the beta potential. Further progress requires derivation of  
$M_{\beta}$. Since $M_{\beta}$ is a constant the present day first Friedmann constraint determines $M_{\beta}$.
\begin{equation}\label{eq-f1}
3\frac{H_0^2}{\kappa^2}=\rho_{\theta_0} + \rho_{m_0}
\end{equation}
where the subscript $0$ denotes the current value.  Using eqn.~\ref{eq-rhop} and~\ref{eq-bdat},  the present day constraint is
\begin{equation}\label{eq-f1co}
3H_0^2=\frac{(H_0\beta(\kappa\theta_0))^2}{2} + M_{\beta}^4((\kappa\theta_0)^2-\delta^2)^2(1-\frac{\beta^2(\kappa\theta_0)}{6}) +\rho_{m_0}
\end{equation}
when rearranged is
\begin{equation}\label{eq-mbc}
3H_0^2(1-\frac{\beta^2(\kappa\theta_0)}{6}) =  M_{\beta}^4((\kappa\theta_0)^2-\delta^2)^2(1-\frac{\beta^2(\kappa\theta_0)}{6}) +\rho_{m_0}
\end{equation}
 Solving for $M_{\beta}$ gives
\begin{equation}\label{eq-m}
M_{\beta}= \sqrt[4]{\frac{3\frac{H_0^2}{\kappa^2}-\frac{\rho_{m_0}}{(1-\frac{\beta^2(\kappa\theta_0)}{6})}}{((\kappa\theta_0)^2-(\kappa\delta)^2)^2}}
\end{equation}
The introduction of the matter density insures that the density in equation~\ref{eq-bv} is the dark energy density, not the
total density.  For completeness the model potential constant $M_{mod}$ is
\begin{equation}\label{eq-mmod}
M_{mod}=\sqrt{\frac{\frac{3H_0^2}{\kappa^2}(1-\frac{\beta^2(\kappa_0)}{6})-\rho_{m_0}}{((\kappa\theta)^2-(\kappa\delta)^2)^2}}
\end{equation}

\subsection{The beta dark energy potential}\label{ss-debpot}
Equation~\ref{eq-pot} shows the three individual polynomial terms of the beta potential $V_{\beta}(a)$.
\begin{equation} \label{eq-pot}
V_{\beta}(\theta) = (M_{\beta}\kappa\theta)^4 -2 (M_{\beta}^2\kappa^2\theta\delta)^2 +(M_{\beta}\kappa\delta)^4)(1-\frac{\beta^2(\kappa\theta)}{6})
\end{equation}
In terms of the Lambert W function the beta potential is
\begin{equation}\label{eq-bpotw}
V_{\beta}(a)=(M_{\beta}\kappa\delta)^4(W(\chi)+1)^2(1-\frac{8W(\chi)}{3(\kappa\delta)^2(W(\chi)+1)^2})=(M_{\beta}\kappa\delta)^4[(W(\chi)+1)^2+\frac{8 W(\chi)}{3(\kappa\delta)^2}]
\end{equation}

\subsection{The Hubble Parameter}\label{ss-H}
Equations~\ref{eq-friedcs} provide the means to derive $H(a)$ and $\dot{H}$ respectively. Similar to the 
calculation of $M_{\beta}$ the first Friedmann constraint yields
\begin{equation}\label{eq-fh}
3H(a)^2(1-\frac{\beta^2(\kappa\theta)}{6}) =  M_{\beta}^4((\kappa\theta)^2-\delta^2)^2(1-\frac{\beta^2(\kappa\theta)}{6}) +\frac{\rho_{m_0}}{a^3}
\end{equation}
The solution for $H(a)$ is
\begin{equation}\label{eq-H}
H(a)=\sqrt{\frac{M_{\beta}^4((\kappa\theta)^2-(\kappa\delta)^2)^2 +\frac{\rho_{m_0}}{a^3(1-\frac{\beta^2(\kappa\theta)}{6})}}{3}}
=\frac{1}{\sqrt{3}}\sqrt{M_{\beta}^4((\kappa\theta)^2-(\kappa\delta)^2)^2 +\frac{\rho_{m_0}}{a^3(1-\frac{8(\kappa\theta^2}
{3((\kappa\theta)^2-(\kappa\theta)^2)^2})}}
\end{equation}
In terms of the Lambert W function
\begin{equation}\label{eq-HW}
H(a)=\sqrt{\frac{(M_{\beta}\delta)^4(W(\chi)+1)^2+\frac{ \rho_{m_0}}{a^3(1+\frac{8W(\chi)}{3(\kappa\delta)^2(W(\chi)+1)^2})}}{3}}
\end{equation}
$H_0$ is included in equation~\ref{eq-HW} via $M_{\beta}$ as given in equation~\ref{eq-m}.

For $\dot{H}$ expanding the right hand side of eqn.~\ref{eq-friedcs} and using eqn.~\ref{eq-pdots} gives
\begin{equation}\label{eq-f2hd1}
3\dot{H} +3H^2=-(\dot{\theta}^2 + \frac{\rho_{m_0}}{a^3} +2p)
\end{equation}
Since $3H^2$ is the sum of the dark energy and matter densities using eqn.\ref{eq-rhop}
for the dark energy pressure gives.
\begin{equation}\label{eq-hdot}
\dot{H}(a)=-\frac{1}{2}(\dot{\theta}^2 + \frac{\rho_{m_0}}{a^3}) =-\frac{1}{2}((\beta H)^2 + \frac{\rho_{m_0}}{a^3})
\end{equation}

\subsection{The derivatives of $\theta$}\label{ss-dott}
Although the time and scale factor derivatives of $\theta(a)$ are not observables they are required for the calculation
of several observable parameters such as the dark energy density.  The $\theta$ derivatives are derived from
equations~\ref{eq-bdat}.  Taking $\dot{\theta}$ as $\beta(\theta)H(\theta)$ yields
\begin{equation}\label{eq-dthetadt}
\dot{\theta}(a)=\frac{4\kappa\theta}{\sqrt{3}}\sqrt{M_{\beta}^4+\frac{\rho_{m_0}}{a^3(((\kappa\theta)^2-(\kappa\delta)^2)^2-\frac{8(\kappa\theta)^2}{3})}}
\end{equation} 
and in terms of the Lambert W function
\begin{equation}\label{eq-dthetadtw}
\dot{\theta}(a)=\frac{4\sqrt{-W(\chi)}}{\sqrt{3}\kappa\delta}\sqrt{(M_{\beta}\kappa\delta)^4+\frac{\rho_{m_0}}{a^3((W(\chi)+1)^2+\frac{8W(\chi)}{3(\kappa\delta)^2})}}
\end{equation}
Also from equations~\ref{eq-bdat} the derivative of $\theta(a)$ with respect to $a$ is $\beta(a)/a$. In the $\theta(a)$ format
\begin{equation}\label{eq-dthda}
\frac{d(\kappa\theta(a)}{da} = =\frac{-4 \kappa\theta}{a(\kappa\theta)^2 -(\kappa\delta)^2 )}
\end{equation}
The Lambert W format is
\begin{equation}\label{eq-dthdaw}
\frac{d(\kappa\theta(a)}{da} = \frac{4\sqrt{-W(\chi)}}{\kappa\delta a(w(\chi)+1)}
\end{equation}
Both the scalar $\theta(a)$ and its derivatives with respect to time and the scale factor are now available enabling
derivation of cosmological parameters and fundamental constants as functions of either the scale factor $a$
or the Lambert $W(\chi(a))$ function.

\subsection{The dark energy density and pressure and the matter density}\label{s-den}
The dark energy and matter densities plus the pressure are important cosmological parameters.  The matter density has 
the simple form
\begin{equation}\label{eq-rhom}
\rho_m(a) = \frac{\rho_{m_0}}{a^3}
\end{equation}
where $\rho_{m_0}$ is the current matter density.  The dark energy density is given by eqns.~\ref{eq-rhop} and~\ref{eq-bdat}.
\begin{equation}\label{eq-rhode}
\rho_{\theta} =\frac{\dot{\theta}^2}{2}+V=\frac{\beta^2H^2}{2}+M_{\beta}^4((\kappa\theta)^2-(\kappa\delta)^2)^2(1-
\frac{\beta^2}{6}) 
\end{equation}
In terms of the scalar $\theta$ the dark energy density is
\begin{equation}\label{eq-rhodeth}
\frac{8(\kappa\theta)^2}{3}(M_{\beta}^4 +\frac{\rho_{m_0}}{a^3(((\kappa\theta)^2-(\kappa\delta)^2)^2-\frac{8(\kappa\theta)^2}{3})})+M_{\beta}((\kappa\theta)^2-(\kappa\delta)^2)^2(1-\frac{8(\kappa\theta)^2}{3((\kappa\theta)^2-(\kappa\delta)^2)^2.})
\end{equation}
and the Lambert W function form is
\begin{equation}\label{eq-rhodew}
-\frac{8W(\chi)}{3(\kappa\delta)^2}[M_{\beta}\kappa\delta)^4 + \frac{\rho_{m_0}}{a^3((W(\chi)+1)^2 +\frac{8W(\chi)}
{3(\kappa\delta)^2})}] +(M_{\beta}\kappa\delta)^4[(W(\chi)+1)^2 +\frac{8W(\chi)}
{3(\kappa\delta)^2}]
\end{equation}

The dark energy pressure is
\begin{equation}\label{eq-pde}
P_{\theta} =\frac{\dot{\theta}^2}{2}-V=\frac{\beta^2H^2}{2}-M_{\beta}^4((\kappa\theta)^2-(\kappa\delta)^2)^2(1-
\frac{\beta^2}{6}) )
\end{equation}
The scalar form of the dark energy pressure is
\begin{equation}\label{eq-pdeth}
P_{\theta} =\frac{8 \theta^2}{3}[M_{\beta}^4 + \frac{\rho_{m_0}}{a^3(((\kappa\theta)^2-(\kappa\delta)^2)^2 - \frac{8 \theta^2}{3})}]
-M_{\beta}^4((\kappa\theta)^2-(\kappa\delta)^2)^2 (1-\frac{8 \theta^2}{3((\kappa\theta)^2-(\kappa\delta)^2)^2})
\end{equation}

The extra term in the dark energy pressure is because the $\frac{8(\kappa\theta)^2}{3(\kappa\delta)^2}M^4$ term 
that canceled in the dark energy density added in the dark energy pressure due to the change in sign of the potential.
Using equation~\ref{eq-bpotw} and~\ref{eq-dthetadtw} the Lambert W forms of the pressure is
\begin{equation}\label{eq-pdew}
P_{\theta}(a)=-(M_{\beta}\kappa\delta)^4(\frac{16W(\chi)}{3(\kappa\delta)^2} + (W(\chi)+1)^2)-\frac{8 \rho_{m_0}W(\chi)}{a^3[3(\kappa\delta)^2(W(\chi)+1)^2+8W(\chi)]}
\end{equation}

\subsection{The Dark Energy Equation of State}\label{ss-wpo}
The dark energy EoS is calculated directly from equation~\ref{eq-deos}. Dividing through by the potential gives
\begin{equation}\label{eq-wv}
w(a)=\frac{\frac{\dot{\theta}(a)^2}{2V(a)}-1}{\frac{\dot{\theta}(a)^2}{2V(a)}+1}
\end{equation}
showing that when $\dot{\theta}(a)^2/2V(a)$ is small $w(a)$ is near minus one.  For most of the cases here 
$\dot{\theta}(a)^2/2V(a)$ does not get much larger than $(w_0+1)$.  The exceptions are the $\delta=3$ cases
at small scale factors. In terms of the scalar $\theta$ $\dot{\theta}(a)^2/2V(a)$ is
\begin{equation}\label{eq-dt2v2}
\frac{\dot{\theta}(a)^2}{2V(a)}=\frac{\frac{8(\kappa\theta)^2}{3}(M_{\beta}^4 + \frac{\rho_{m_0}}{a^3(((\kappa\theta)^2-(\kappa\delta)^2)^2-\frac{8(\kappa\theta)^2}{3})})}{M_{\beta}^4((\kappa\theta)^2-(\kappa\delta)^2)^2(1-\frac{8(\kappa\theta)^2}{3((\kappa\theta)^2-(\kappa\delta)^2)^2})}
\end{equation}
and
\begin{equation}\label{eq-wtheta}
w(\kappa\theta)=\frac{\frac{8(\kappa\theta)^2}{3}(M_{\beta}^4 + \frac{\rho_{m_0}}{a^3(((\kappa\theta)^2-(\kappa\delta)^2)^2-\frac{8(\kappa\theta)^2}{3})})-{M_{\beta}^4((\kappa\theta)^2-(\kappa\delta)^2)^2(1-\frac{8(\kappa\theta)^2}{3((\kappa\theta)^2-(\kappa\delta)^2)^2})}}{\frac{8(\kappa\theta)^2}{3}(M_{\beta}^4 + \frac{\rho_{m_0}}{a^3(((\kappa\theta)^2-(\kappa\delta)^2)^2-\frac{8(\kappa\theta)^2}{3})})+{M_{\beta}^4((\kappa\theta)^2-(\kappa\delta)^2)^2(1-\frac{8(\kappa\theta)^2}{3((\kappa\theta)^2-(\kappa\delta)^2)^2})}}
\end{equation}
Using equations~\ref{eq-rhodew} and~\ref{eq-pdew} the Lambert W equation for $w(\chi)$ is
\begin{equation}\label{eq-ww}
w(\chi)=\frac{-M_{\beta}^4(\frac{16W(\chi)}{3(\kappa\delta)^2}+(W(\chi)+1)^2)-\frac{8\rho_{m_0}W(\chi)}{a^3(3(\kappa\delta)^2(W(\chi)+1)^2+8W(\chi))}}{M_{\beta}^4(W(\chi)+1)^2-\frac{8\rho_{m_0}W(\chi)}{a^3(3(\kappa\delta)^2(W(\chi)+1)^2+8W(\chi))}}
\end{equation}

\subsection{The CPS templates}\label{ss-cpst}
The functions for the cosmological parameters derived in this section form the set of CPS templates for flat quintessence
and the HI potential.    Appendix~\ref{as-A} gathers the templates together in an abbreviated form, giving the templates both
as functions of $\theta(a)$ to display the physics and as functions of the Lambert W functions $W(\chi)$ , where $\chi =
q a^{\frac{8}{\delta^2}}$, to aid in template coding.  The Lambert W function is a standard function in most mathematical coding
platforms.  The following section displays the evolutions of the observable  and some of the unobservable cosmological
parameters such as the scalar $\theta$ that are essential elements in the cosmology.  Section~\ref{s-fc} follows with the 
evolution of the fundamental constants $\alpha$ and $\mu$.

\section{Cosmological Parameter Evolution and Observations}\label{s-cpeo}
The HI potential provides a rich set of evolutionary characteristics that are not obvious from their mathematical form in the 
templates. This section provides graphical presentations of the evolutions and comments on the observational consequences 
of the parameters for discriminating between static and dynamical dark energy.  The evolutions are a strong function of the 
boundary conditions and the selection of relevant constants such as the HI potential constant $\delta$.

\subsection{Boundary conditions and constants}\label{ss-bcc}
The boundary conditions and values of relevant constants are displayed in table~\ref{tab-ival}.  The boundary conditions are
set at the current epoch and the constants are chosen to display the range of evolutions.
\begin{table}
\caption{ \label{tab-ival} Boundary conditions and parameter values in this work. $H_0$ is the current value of the
Hubble parameter in units of $\frac{km/sec}{Mpc}$. $\Omega_{m_0}$ and $\Omega_{\phi_0}$ are the current ratios 
of the matter and dark energy densities to the critical density. $w_0$ are the current values of the dark
energy equation of state and $\kappa\delta$ are the dimensionless constant values in the dark energy potential.}
\begin{tabular}{|c|c|c|ccc|ccc|c|}
\hline
$H_0$ &$\Omega_{m_0} $&$\Omega_{\theta_0}$& &$ w_0$ & & &$\kappa\delta$& & scale factor $a$\\
\hline
70 & 0.3& 0.7 &-0.99&-0.995&-0.999&1&2&3&0.1 - 1.0\\
\hline
\end{tabular}
\end{table}
The form of the evolutions is primarily controlled by two parameters, the current dark energy EoS, $w_0$ and the HI
potential constant $\delta$.  The seemingly arbitrary choices for $\delta$, 1, 2, and 3, display
the three different characteristics of HI potential evolutions.  Since discrimination between static and dynamical dark
energy is a main subject of the study the $w_0$ values are placed close to minus one to show the difficulties
of such discrimination.  The choices for $H_0$ and $\Omega_{\theta}$ are set to be close matches to 
concordance cosmology values.  All are easily changed in the templates to suit other boundary conditions and constant
choices.

\subsection{The evolution of the scalar}\label{ss-te}
\begin{figure}
\scalebox{.9}{\includegraphics{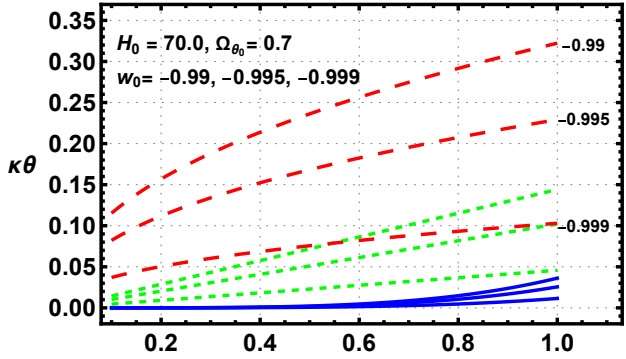}}
\caption{The evolution of the scalar $\kappa\theta(a)$ for all $\kappa\delta$ and $w_0$ values. The $\delta=1$ evolutions are  
solid tracks, blue, $\delta=2$ the dashed tracks, green and $\delta=3$ the long dashed tracks, red.  These same line styles 
and colors are used to denote $\delta$ in the remaining figures unless stated otherwise. The $w_0$ values are 
indicated on the right for the $\delta = 3$ tracks.  For all $\delta$ values the $w_0 = -0.99$ case has the greatest evolution and 
the $w_0 = -0.999$ case has the least.}
\label{fig-t12e}
\end{figure}

The evolution of $\theta$ depends heavily on  $w_0$ and $\delta$.  Figure~\ref{fig-t12e}  shows the evolution of the scalar 
for all combinations of $w_0$ and $\kappa\delta$ as a function of $a$.  The $\delta =1$ evolution traces are solid lines, blue, 
$\delta=2$ dashed lines, green, and $\delta=3$ long dashed lines, red,. This format is retained in subsequent figures unless 
specified otherwise.
The degree of evolution increases with larger values of $\delta$ and further deviation of $w_0$ from minus one.  The second 
derivative of the slope changes from positive for $\delta=1$, to almost zero for $\delta=2$ to negative for $\delta=3$. This
change in second derivative persists for most evolutions and results in non-monotonic evolution for the dark energy EoS
not seen in previous monotonic potential calculations \cite{thm18, thm19}. The relative evolution of the scalar is quite small, 
particularly for the $\delta = 1$ cases, consistent with a small value of $\dot{\theta}$. The values in fig.~\ref{fig-t12e} are for 
$\kappa\theta$ which is not the true scalar.  The true scalar is $M_{\beta}\kappa\theta$ where $M_{\beta}^4$ is the leading 
dimensional constant in the potential. Each combination of $\delta$ and $w_0$ requires a different value of $M$ to meet all 
of the boundary conditions as shown by equation~\ref{eq-m}.  

\subsection{The evolution of the dark energy potential}\label{ss-evde}
\begin{figure}
\scalebox{0.79}{\includegraphics{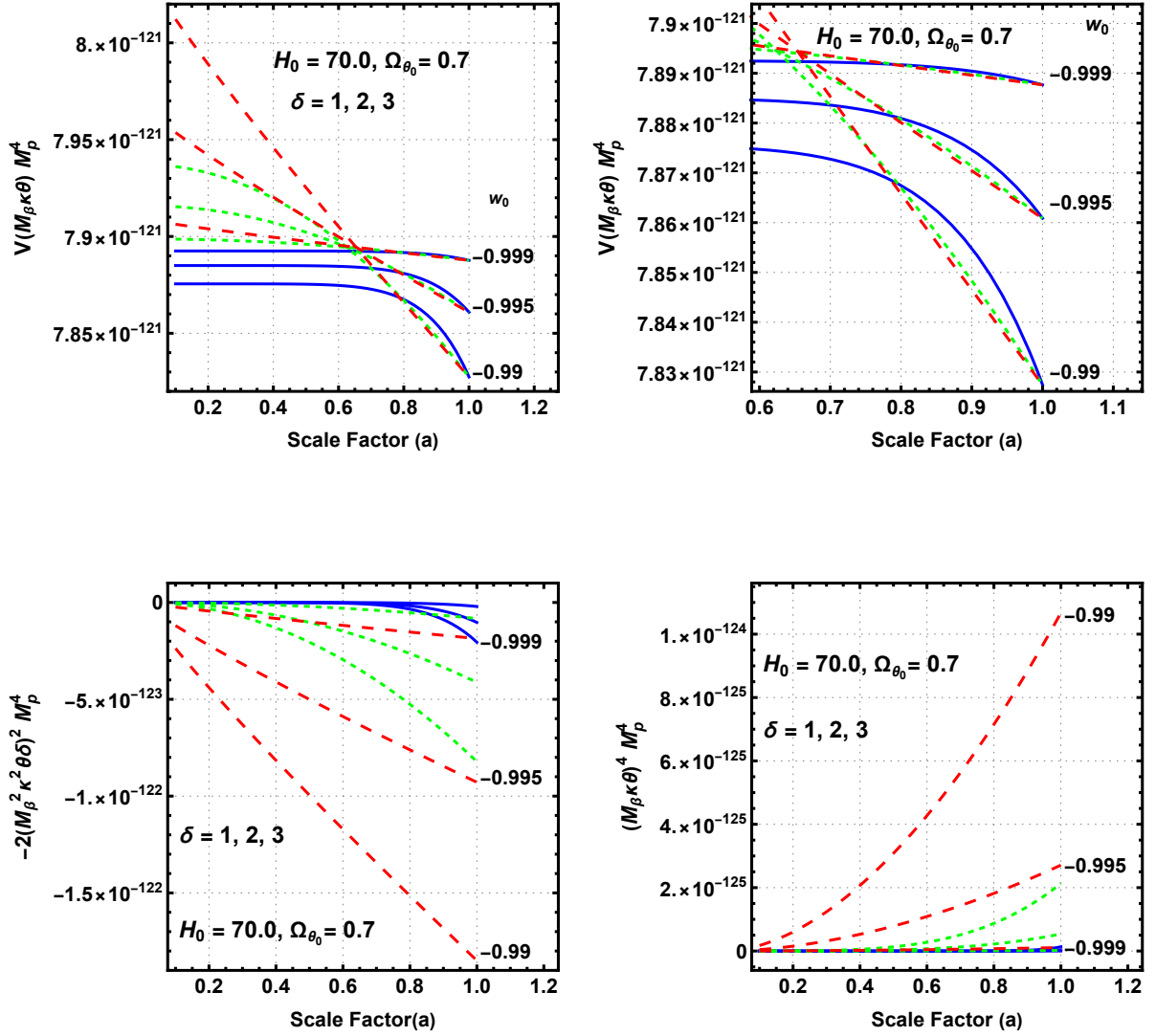}}
\caption{The top left panel shows the evolution of $V(a)$ for all values of $w_0$ and $\delta$.The values of $w_0$ are shown at
 the bottom right of the figures and the line styles are consistent with previous plots. The top right panel shows the late time 
 evolution of the potential  The bottom left plot shows the evolution of the first variable term $-2(\kappa^2 M_{\beta}^2 \theta 
 \delta)^2$  and the bottom right plot is the second variable term $(M_{\beta} \kappa \theta)^4$. Note the changes in vertical
 scale in the bottom two plots.}
\label{fig-evde}
\end{figure}

Figure~\ref{fig-evde} shows the monotonically decreasing evolution of $V(a)$ for all values of $\delta$ and $w_0$.   It also 
plots the evolution of the two dynamical terms $-2(M_{\beta}^2\kappa^2\theta \delta)^2$ and $(M_{\beta} \kappa \theta)^4$.  
The potential has very small changes over the range of scale factors.  Though not required in the beta function formalism the 
potential conforms to the usual slow roll conditions.
The polynomial form of the potential produces important differences from monomial potentials.  The constant $(M_{\beta}\kappa
\delta)^4$ is the dominant term since $\theta \ll \delta$ at the $a$ values considered here.  It is similar to a cosmological 
constant. The two dynamical terms are  $-2(M_{\beta}^2\kappa^2\delta\theta)^2$ and $(M_{\beta}\kappa\theta)^4$  with the 
first term dominant over the second again due to $\delta > \theta$.  For the $\delta=1$ case the potential is essentially 
constant until the late time evolution considered in this work.

\subsection{The accuracy of the CPS templates}\label{ss-acpst}
\begin{figure}
\scalebox{.65}{\includegraphics{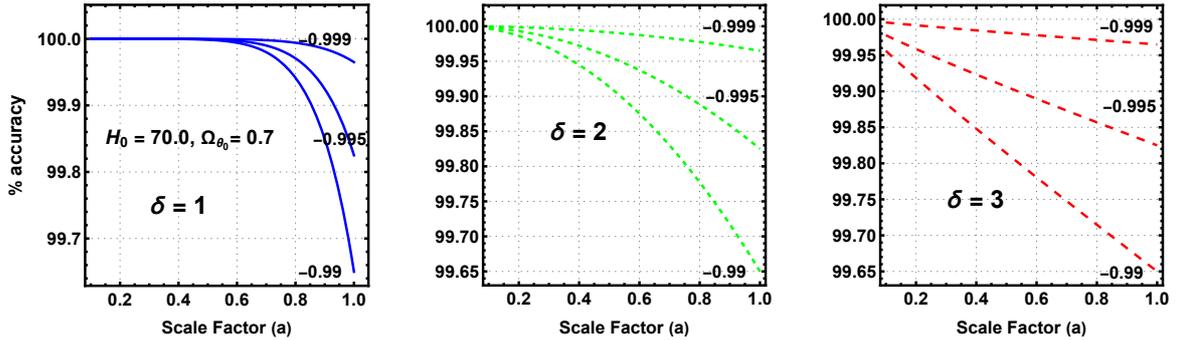}}
\caption{The figure shows the percentage accuracy that the beta potential represents the model potential for each value
of $\delta$.}
\label{fig-ac}
\end{figure}

The CPS templates are exact and consistent solutions for the beta potential and the flat quintessence cosmology.  The
CPS templates are therefore accurate for the model potential to the degree that the beta potential is accurate for the
model potential.  The tradeoff of the beta potential for the model potential is deemed acceptable since it enabled the
derivation of the scalar as an analytic function, the Lambert W function, in terms of the scale factor.  No analytic beta
function was found that produced the model potential in equation~\ref{eq-bvbc}.  The beta potential is simply
$(1-\beta^2/6)$ times the model potential therefore the accuracy is $(1-\beta^2/6)$. Figure~\ref{fig-ac} shows the percentage 
accuracy that the beta potential represents the model potential for all cases. 

The maximum deviation of the beta potential from the model potential is at the current epoch.  The deviations for
a given $w_0$ are all the same since the beta functions are all the same for a given $w_0$. The $a=1$ percentage 
accuracies are 99.65, 99.825 and 99.965 for $w_0$ equal to -0.99, -0.995 and -0.999 with the accuracies for scale
factor less than one all higher.   The paths to the 
common $a=1$ accuracy differ for each $\delta$.  The accuracy of the $\delta=1$ case is essentially $100 \%$
up to a scale factor of 0.6 before evolving rapidly to the final accuracy at $a=1$.  The $\delta =2$ cases deviate
from $100\%$ very early and evolve with increasing rapidity to the final value.  The $\delta=3$ cases begin deviating 
from $100\%$ at scale factors less than 0.1 and evolve linearly to the final values.  The cosmological parameters
set by the boundary conditions, however, are exactly correct at $a=1$.  

\subsection{The evolution of the Hubble parameter}\label{ss-He}
\begin{figure}
\scalebox{.9}{\includegraphics{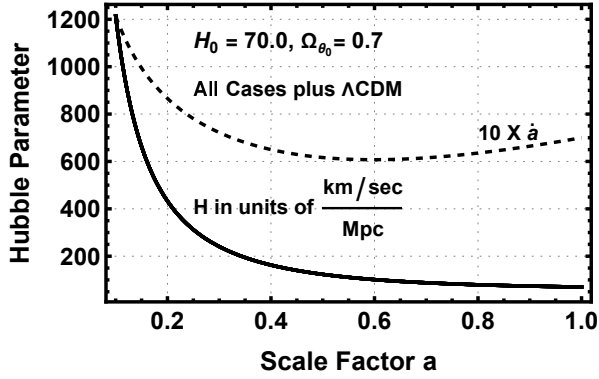}}
\caption{The evolution of $H(a)$ for all values of $w_0$ and $\delta$ plus $\Lambda$CDM in units of (km/sec)/Mpc.
The traces are indistinguishable at the resolution of the figure. The dashed track shows the evolution of $10\times \dot{a}$
to indicate the transition from deceleration to acceleration of the expansion of the universe. The $\dot{a}$ track is 
multiplied by 10 to make the transition visible at the scale of the plot.}
\label{fig-hall}
\end{figure}

Figure~\ref{fig-hall} shows the evolution of the Hubble parameter $H(a)$ for all values of $\delta$ and $w_0$
plus $\Lambda$CDM.  As found for other dark energy potentials \citep{thm18, thm19} the individual traces are 
indistinguishable at the resolution of the figure making the Hubble parameter a poor discriminator between 
static and dynamical dark energy. This indicates that the Hubble parameter is relatively insensitive to the dark energy
potential, unlike most of the other parameters.  This is partially due to fixing $H$ at $a=1$ to $H_0$ and the dominance
of matter over dark energy at redshifts greater than one. The deviations from $\Lambda$CDM are quantified in 
section~\ref{ss-hdev}
The dashed line in the figure shows the evolution of $\dot{a}$ confirming that the transition from deceleration to
acceleration of the expansion of the universe occurs at the observed epoch.

\subsubsection{Deviations from $\Lambda$CDM}\label{ss-hdev}
\begin{figure}
\scalebox{.53}{\includegraphics{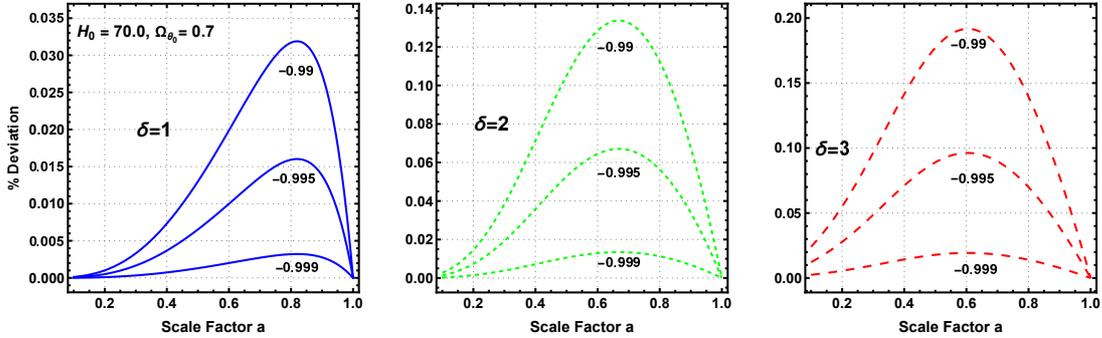}}
\caption{The evolution of the percentage difference between the HI Hubble parameter and $\Lambda$CDM for all values of 
$w_0$ and $\delta$. Each of the three panels is for a specific $\delta$.}
\label{fig-hdf}
\end{figure}

The percentage deviation of the dynamical dark energy Hubble parameter from $\Lambda$CDM, $\frac{H_{dyn}-H_{\Lambda}}
{H_{\Lambda}}$, is a quantitative measure of the difference between the evolutions. Figure~\ref{fig-hdf} shows the percentage
deviations for all of the cases as a function of $a$.
All deviations are less than $0.19\%$ at their maximum and  all of the $\delta =1$ deviations are less than $0.033\%$.
The maximum deviation from $\Lambda$CDM for the $\delta=1$, $w_0=-0.999$ case is less than  $0.0033\%$. This is
not detectable with the current accuracy of $H(a)$ measurements.  Maximum deviations occur at smaller $a$ as the value 
of $\delta$ increases. but all of the cases have maximum deviations at $a$  greater than 0.5.  For a given $H_0$ the 
deviation is zero at redshift zero by the boundary conditions and increases at higher redshifts. After a redshift of $\approx1$ 
matter density begins to dominate, decreasing the percentage deviation.

\subsubsection{Hubble parameter observational implications}\label{sss-hpoi}
For the $w_0$ values close to minus one in this study it is difficult to detect deviations from $\Lambda$CDM using only 
Hubble parameter observations.  The peak deviations occur between scale factors of 0.6 and 0.8 dependent on the $\delta$ 
value and essentially independent of $w_0$.  This low redshift regions offers the best chance for observing departures from 
$\Lambda$CDM, particularly for higher deviations of $w_0$ from minus one than in this work at high $\delta$ values.  
Bringing $w_0$ closer to minus one reduces the deviations roughly as $(w+1)$. This means that observational constraints 
on deviations of $H$ from $\Lambda$CDM can be met by bringing $w_0$ closer to minus one while still  maintaining a dynamical dark energy density.  Falsifying dynamical dark energy with the Hubble parameter alone is very  difficult.  The 
$\delta = 1$, $w_0=-0.999$ deviations are undetectable with current techniques.  Peculiar motions at low redshift may be a 
complicating factor at the highest sensitive scale factors.  The $\delta=3$, $w_0=-0.99$ deviations at $0.2 \%$ are most likely 
within reach with future observational  techniques given that the current accuracy is on the order of 1km/sec \cite{rie21}.

\subsection{Evolution of $\dot{\theta}$}
\begin{figure}
\scalebox{.7}{\includegraphics{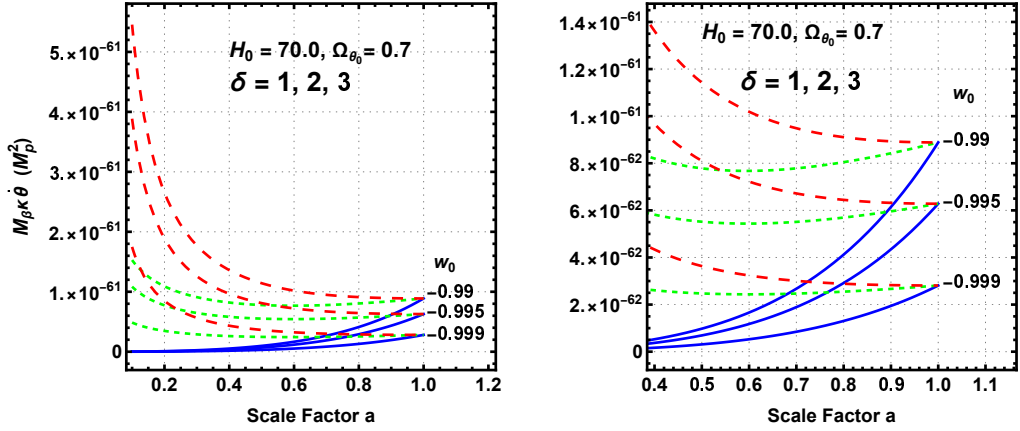}}
\caption{The evolution of the time derivative true scalar $M_{\beta} \kappa \dot{\theta}$ as a function of the scale factor for all 
values of $\delta$ and $w_0$ for the full scale factor range in the upper panel and in more detail for $a$ between 0.4 and 1
.0 in the lower panel.}
\label{fig-tdot}
\end{figure}

Figure~\ref{fig-tdot} shows the evolution of $M_{\beta}\kappa\dot{\theta}$ for the full range of $a$ and for $a$ between 0.4 
and 1.0 to show the late time evolution in more detail.
It is obvious that the $a=1$ $\dot{\theta}$ values for a specific $w_0$ are equal because $H=H_0$ and the $a=1$ $\beta$
values are the same for a given $w_0$. The first derivative of the evolution changes from positive for $\delta=1$, transitioning 
from negative to positive for $\delta=2$  and negative for $\delta=3$. These changes are a primary driver for the changes in the 
second derivative in many of the cosmological parameters.  This is particularly true for the dark energy EoS.

\subsection{Evolution of the dark energy density}\label{ss-dede}
\begin{figure}
\scalebox{.53}{\includegraphics{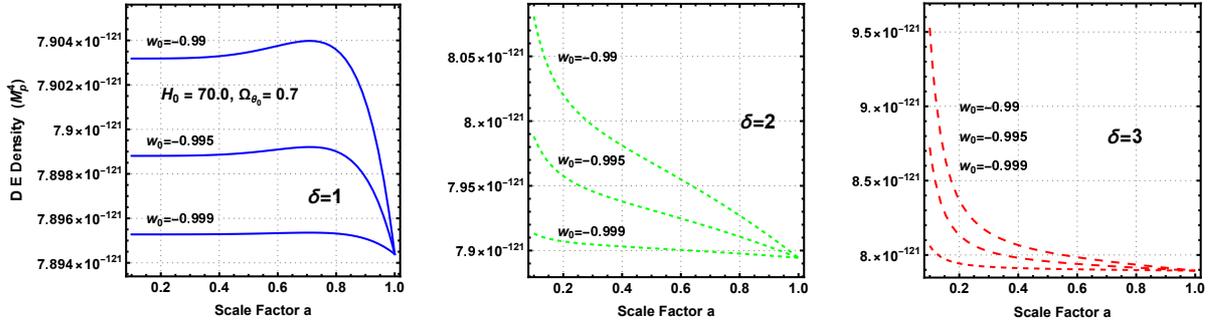}}
\caption{The evolution of the dark energy density $\rho_{\theta}$ as a function of the scale factor.  The panels are for
$\delta$ =1, 2, and 3 left to right.}
\label{fig-deden}
\end{figure}

Figure~\ref{fig-deden} shows the evolution of the dark energy density for all of the cases.
The evolutions are sensitive to the value of $\delta$.  The $\delta=1$ evolutions are quite flat for scale factors less
than 0.5 with a slight non-monotonic rise before decreasing with a positive second derivative to the common point at $a=1.0$.
The $\delta=2$ evolution is monotonically decreasing initially with a negative second derivative transitioning to a slightly
positive second derivative near $a=0.5$.  The $\delta=3$ evolution is monotonically decreasing  with a negative second
derivative at all plotted scale factors.  The total change in the dark energy density is less than $0.1\%$ for all of the $\delta=1$
cases, on the order of $1\%$ for $\delta = 2$ and on the order of $10\%$ for $\delta=3$.  As usual the closer $w_0$ is
to minus one the less the evolution.  The small percentage change of $\rho_{\theta}$ is consistent with the cases having
evolutions similar to $\Lambda$CDM.  The flat $\rho_{\theta}$ regions in the $\delta=1$ cases act like  $\Lambda$CDM.

\begin{figure}
\scalebox{.9}{\includegraphics{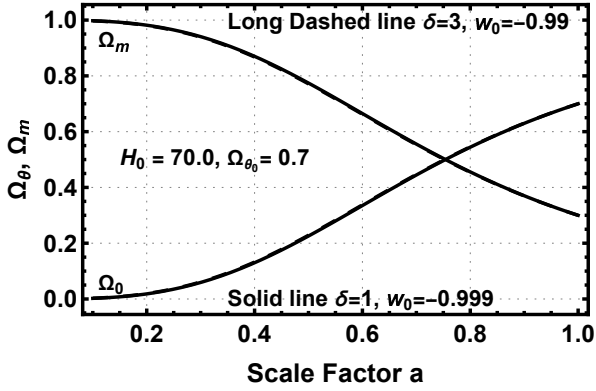}}
\caption{The evolution of $\Omega_{\theta}$ and $\Omega_m$ for  $\delta=3$, $w_0=-0.99$, long dashed line and $\delta=1$, 
$w_0=-0.999$ solid line. The tracks are indistinguishable at the resolution of the plot.}
\label{fig-omega}
\end{figure}

Figure~\ref{fig-omega} shows the evolution of $\Omega_m$ and $\Omega_{\theta}$ for the two most extreme input values, 
$\delta=3$, $w_0=-0.99$, long dashed line and $\delta=1$, $w_0=-0.999$ solid line.  The tracks are indistinguishable 
in the plot indicating they are relatively independent of $\delta$ and $w_0$.

\subsection{Evolution of the dark energy equation of state}\label{ss-we}
\begin{figure}
\scalebox{.53}{\includegraphics{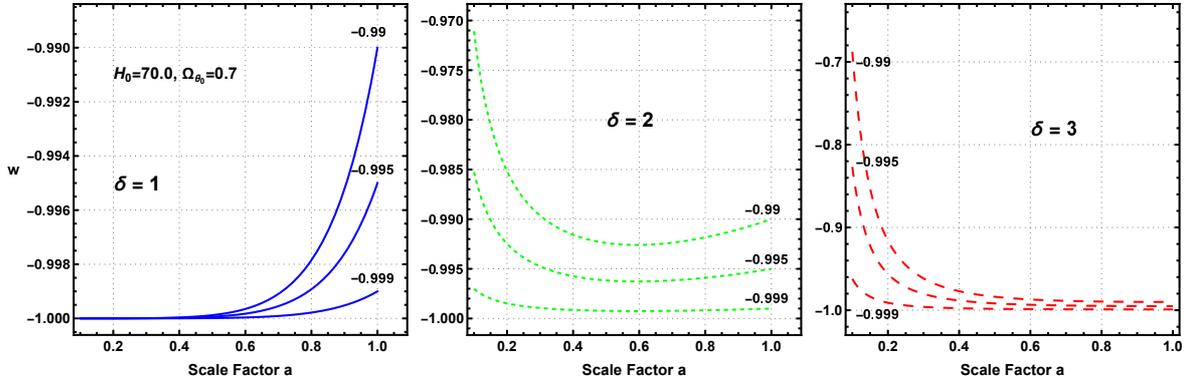}}
\caption{The evolution of $w$ as a function of the scale factor $a$. The left to right panels are for $\delta$ = 1, 2, and 3.  
The $w_0$ values are indicated by numbers near the traces.}
\label{fig-w}
\end{figure}

The evolution of the dark energy EoS $w$ is very dependent on the value of $\delta$ as shown in figure~\ref{fig-w}. The 
left panel of figure~\ref{fig-w} shows the evolution for the $\delta=1$ cases.
The $\delta=1$ cases are purely thawing at all scale factors in figure~\ref{fig-w}.  Since they are thawing $w_0$ is its 
maximum deviation from minus one. For $a<0.6$ $w$ is essentially minus one.  These are the cases most similar to 
$\Lambda$CDM and the hardest to distinguish from a cosmological constant.  

The $\delta=2$ cases in the middle panel transition from freezing to thawing at roughly $a=0.5$, giving a non-monotonic 
evolution of $w$. Their early time $w$ values are higher than their $w_0$ values but their overall evolutions are almost as 
flat as the $\delta=1$ thawing cases. The right hand panel $\delta=3$ cases are classic freezing evolutions with $w$ initially 
distant from minus one freezing toward minus one.  Detection of their high deviations from minus one may be possible with 
the upcoming more precise observational measurements.  Identification of the $\delta=3$ evolutions would be a strong case 
in favor of dynamical dark energy.  The large deviations from minus one, $\approx -0.7$ at high redshift, offer the best chance of
observational detection.

The $\delta=1$ $w$ evolutions illustrate the difficulty of falsifying dynamical dark energy through dark energy EoS
observations.  All of the $\delta=1$ cases have deviations from minus one that are too small to be detected by
current observational techniques.  Any non-zero constraint on deviations of $w$ from minus one can be met by 
simply making $w_0$ closer to minus one but not exactly minus one  On the other hand confirmation of a dynamical
dark energy by observations establishing values of $w$ not equal to minus one is strong evidence for a dynamical
dark energy density.  Another possibility is a detection of a temporal variation of fundamental constants such as the fine 
structure constant $\alpha$ or the proton to electron mass ratio $\mu$ discussed in section~\ref{s-fc}.

Figure~\ref{fig-w} displays the versatility of the HI potential.  The simple $\delta$ = 1, 2, 3 values were picked because
they demonstrate the three types of evolution: pure thawing, transition from freezing to thawing and freezing.  Intermediate
values of $\delta$ provide smooth transitions between the three modes. This means that the HI potential CPS templates can
easily cover the a large range of dark energy EoS evolutions.

\section{Temporal evolution of fundamental constants}\label{s-fc}
In the absence of special symmetries scalar fields that interact with gravity also interact with other sectors \citep{car98, fig19}.
The proton to electron mass ratio $\mu$ and the fine structure constant $\alpha$ values are determined by the Quantum 
Chromodynamic Scale $\Lambda_{QCD}$, the Higgs vacuum expectation value $\nu$ and the Yukawa couplings
$h$ \citep{coc07}. Here it is assumed that the scalar responsible for dark energy also interacts with these sectors.  The total 
of the interactions with the three particle physics parameters produces a coupling factor $\zeta_x$, where $x$ is
either $\mu$ or $\alpha$  \citep{coc07, thm17}.  Since the couplings are due to the same scalar field there is a relationship  
between the evolution of $w$ and the evolution of $\mu$ and $\alpha$ described in section~\ref{ss-fcc}.

The relative change of the constants is given by eqn.~\ref{eq-dfc}.
\begin{equation} \label{eq-dfc}
\frac{\Delta x}{x}=\zeta_x (\kappa\theta-\kappa\theta_0)
\end{equation}
This simple linear interaction can be thought of as the first term of a Taylor expansion of a more complex coupling.
The only difference in the evolution of the constants is the strengths of $\zeta_{\mu}$  and $\zeta_{\alpha}$
The current limits on the variance of $\alpha$ and $\mu$ are  $\Delta \alpha / \alpha=-(1.3 \pm1.3_{stat} \pm0.4_{sys})
\times 10^{-6}$,  $1\sigma$ \citep{mur21} at $z=1.15$ and  $\Delta \mu / \mu \leq \pm 1.1\times 10^{-7}$ $2\sigma$ 
\citep{bag13,kan15} at $z= 0.89$. Both redshifts have look back times greater than half the age of the universe.  The following
analysis uses the more stringent limit on the variation of $\mu$.

\begin{figure}
\scalebox{0.6}{\includegraphics{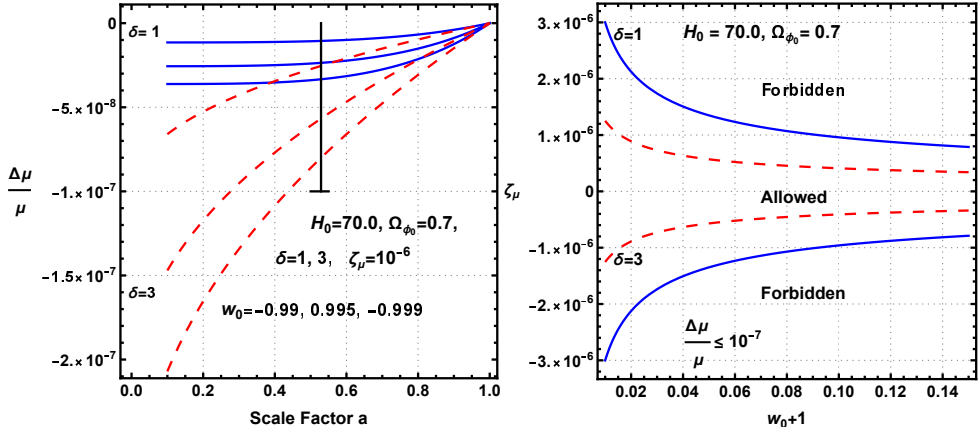}}
\caption{\label{fig-fce} The left plot shows evolution of $\frac{\Delta \mu}{\mu}$ for the three $w_0$ values and $\delta$ 
values of 1 and 3. The $\delta=2$ evolution lies between the two but is not shown to avoid crowding.  The error bar at 
$a=0.53$ extending from 0 to $-1.1\times10^{-7}$ is the $2\sigma$ observational constraint on $\frac{\Delta \mu}{\mu}$. 
The right plot shows the forbidden and allowed regions in the $(w_0+1)$, $\zeta_{\mu}$ plane are shown for $\delta$ values 
of 1 (solid blue lines) and 3 (red  long dashed lines,  The regions between the positive and negative tracks for each 
$\delta$ value are allowed while the regions outside of the two tracks are forbidden.}
\end{figure}

The left panel of figure~\ref{fig-fce} shows the evolution of $\Delta \mu/\mu$ as a function of the $a$ for $\zeta_{\mu}=10^{-6}$.  
The error bar at $a= 0.53$ is the $2\sigma$ observational $\Delta \mu / \mu$ constraint given above.  All of the cases
considered in this work satisfy the $2\sigma$ $\Delta\mu/\mu$ constraint, mainly due to the small deviation of $w_0$
from minus one.  For larger deviations from minus one the $\delta=3$ cases start to violate the constraint.  The
$\delta=1$ cases are close to the constraint and improved measurements could produce constraints not satisfied by
even the $\delta=1$ cases.  These cases, however, can be brought back into compliance by reducing the particle 
physics coupling factor as covered in section~\ref{ss-fcc}.  The coupling strength was chosen to be positive making the 
evolution negative but the opposite is equally likely. 
Consistent with equation~\ref{eq-dfc} the second derivative of the evolution of $\mu$ is the same as the scalar $\theta$. 

\subsection{Fundamental constant cosmological and new physics constraints}\label{ss-fcc}
The evolution of the fundamental constants in a quintessence cosmology has a direct relationship to the
evolution of  the dark energy EoS.  The evolution of $\Delta \mu/\mu$  is \cite{thm13}.
\begin{equation}\label{eq-fcw}
\frac{\Delta \mu}{\mu} = \zeta_{\mu}\int_1^a \sqrt{3\Omega_{\theta}(x)(w(x)+1)}x^{-1}dx
\end{equation}
Equation~\ref{eq-fcw} shows that  $\mu$ or $\alpha$ varies whenever $w$ is different than  minus one even if the scalar is 
not varying with time.  This makes $\alpha$ and $\mu$ $w$ meters in the universe.  The constraints on the temporal variation of 
the fundamental constants are constraints on the combination of the cosmological parameter $w$ and the new physics coupling 
constant $\zeta_x$.  The fundamental constant constraints  therefore  define forbidden and allowed regions in a particle physics - 
cosmological plane defined by $\zeta_x$ and $w$.  The lack of a detected variance is consistent with but not a confirmation
of $\Lambda$CDM.  A detection of any variance of the fundamental constants, however, would signal the existence of new physics.

Making $w_0$ closer to minus one is a cosmological adjustment, lowering the value of $\zeta_{\mu}$ is a new physics 
adjustment, setting $w_0$ to minus one and $\zeta_{\mu}$ to zero is the standard model. The right panel of  figure~\ref{fig-fce} 
shows the forbidden and allowed zones for  $\delta$ = 1 and 3. It shows that there is significant space in the $(w_0 +1)$ - 
$\zeta_{\mu}$ plane to satisfy the observational constraints with a dynamical cosmology  The tracks between the allowed 
and forbidden zones are defined by
\begin{equation}\label{eq-fcp}
\zeta_{\mu} = \frac{\pm\frac{\Delta \mu}{\mu}}{\kappa\theta(a_{ob},w_0,\Omega_{\theta_0})-\kappa\theta_o(1,w_0,\Omega_{\theta_0})}
\end{equation}
where $a_{ob}$ is the $a$ of the observation. The terms $w_0$ and $\Omega_{\theta_0}$ appear due to their
presence in eqn.~\ref{eq-to} for $\theta_0$.  The $\delta = 2$ track lies between tracks for $\delta = 1$ and 3.  The higher 
$\delta$ values have greater $\Delta \theta$ and therefore tighter constraints.  At the smallest plotted value of $(w+1)=0.01$ 
the $\zeta_{\mu}$ magnitude for $\delta=3$ is constrained to $|\zeta_{\mu}| < \pm1.22\times10^{-6}$.  The constraint on
$\Delta\mu/\mu$ puts constraints on the temporal variation of the particle physics parameters of $\Delta\Lambda_{QCD}/
\Lambda_{QCD} \le \pm 7.9 \times 10^{-5}$, $\Delta\nu/\nu \le 7.9\times10^{-5}$ and $\Delta h/h \le 4.9\times10^{-7}$
\cite{thm17}. 

\section{Conclusions}\label{s-con}
The beta function formalism provides a methodology for producing accurate evolutionary templates for a model dark energy
potential using a beta function potential that is an accurate but not exact duplicate of the model potential.  This methodology
produced Cosmology and Potential Specific, CPS, templates for a flat quintessence cosmology with a model dark energy
potential having the same mathematical form as the Higgs potential, the HI potential.  CPS templates for the Hubble Parameter,
the dark energy and matter densities, and the dark energy EoS were calculated for nine different combinations of the current
dark energy EoS, $w_0$ and the HI potential constant $\delta$ as analytic functions of the scale factor $a$.  These templates 
offer significantly improved accuracies relative to the parameterizations commonly used in likelihood calculation for dynamical
cosmologies.  They avoid the possibility of large errors in cosmology likelihoods by replacing parameterizations with specific
cosmology and dark energy potential physics based templates.  The new physics based CPS templates are good candidates
for fiducial templates to discriminate between static and dynamical dark energy using the current and expected future 
cosmological parameter databases.   With this purpose in mind the templates calculated in this work have current dark
energy EoS values close to minus one to produce dynamical evolutions similar to the $\Lambda$CDM evolutions.  Given
the analytic functions presented in the text and assembled in appendix~\ref{as-A} CPS templates for a different set of
boundary conditions are easily created.  The methodology is also adaptable to the particular physics of other cosmologies and 
dark energy potentials.

The polynomial HI potential produces a wide range of cosmological parameter evolutions.  A common feature of the evolutions 
are transitions of the sign of the second derivative of the evolution between negative and positive as a function of the
constant $\delta$ in the HI potential.  For the dark energy EoS $w$ the second derivative is positive for $\delta=1$, thawing,
transitioning from negative to positive for $\delta=2$, freezing to thawing, and negative for $\delta=3$, freezing, producing a 
comprehensive range of $w$ evolutions.  At early times the constant term in the potential dominates, acting like a cosmological 
constant.

The evolution of the Hubble parameter is very similar to the $\Lambda$CDM evolution partly due to the choice of $w_0$ values
close to minus one and partly that the Hubble parameter is relatively insensitive to the dark energy potential in a quintessence
cosmology \cite{thm18,thm19}.  This highlights the difficulty of using only the Hubble parameter to discriminate between static 
and dynamical dark energy density.  The thawing $w$ evolutions for the $\delta=1$ cases are similarly difficult to discriminate 
from $\Lambda$CDM since $w$ is initially at minus one and only evolves away from minus one at late times to meet their
$w_0$ almost minus one boundary conditions. Appendix~\ref{as-A} contains an abridged version of a CPS template suite
for flat quintessence with the HI dark energy potential.

The $\delta=1$ CPS templates, particularly with $w_0=-0.999$ are excellent checks on whether dynamical cosmologies can 
achieve the same likelihoods as $\Lambda$CDM.  This property makes these CPS templates good candidates for standard
tests in all likelihood calculations that involve discrimination between static and dynamical dark energy.  Lower likelihoods
for these templates relative to $\Lambda$CDM can be due to a lower likelihood of quintessence and the HI potential as
opposed to evidence for static dark energy.

The cosmological and particle physics constraints from the observed limits on the temporal evolution of the 
fundamental constant $\alpha$ and $\mu$ are also calculated.  All cases studied here met these limits for the most
stringent limit on $\Delta\mu/\mu$ with a coupling constant of $10^{-6}$.  All of the $\delta=3$ cases, however, would have 
failed if the coupling constant was a factor of two higher.  The constraint on $\Delta\mu/\mu$ imposes both a cosmological
constraint on $w$ and a particle physics constraint on temporal evolution on the Quantum Chromodynamic Scale 
$\Lambda_{QCD}$, the Higgs vacuum expectation value $\nu$ and the Yukawa couplings $h$ that comprise the coupling
constant $\zeta_{\mu}$.  These define an allowed area in the $\zeta_{\mu}$ - ($w_0+1)$ plane that constrains both
cosmology and particle physics.  The allowed regions shown in Figure~\ref{fig-fce} include $\Lambda$CDM and
the standard model at the 0.,0. origin but also retain significant space for a dynamical dark energy as well.

\acknowledgments
The author would like to acknowledge the very useful and informative discussion with Sergey Cherkis on the 
mathematical properties of the Lambert W function.

\appendix 
\section{CPS templates in $W(\chi)$}\label{as-A}
$W(\chi)$, the solution to $\chi = W(\chi)e^{W(\chi)}$, is a central function for quintessence with the HI potential.  The variable 
$\chi =-\frac{a^{\frac{8}{(\kappa\delta)^2}}}{(\kappa\delta)^2}e^{\frac{c}{(\kappa\delta)^2}}$. It can be written with three 
parameters: the scale factor $a$, the constant $\delta$ and a compound constant $q=-\frac{e^{\frac{c}{(\kappa\delta)^2}}}
{(\kappa\delta)^2}$ where $c$ is defined in the main text and given below.. The boundary conditions are in table~\ref{tab-ival}.
The dimensionless scalar is $\kappa\theta$ and the subscript 0 indicates the value at the present time. In the following the
first function is the parameter or constant in terms of $\theta$ and the second is in terms of $W(\chi)$.\vspace{0.2cm}

\noindent
\textbf{The scalar field $\mathbf{\kappa\theta}$} \vspace{0.15cm}

\noindent
$\kappa\theta$, \hspace{1cm} $\kappa\delta\sqrt{-W(\chi)}$. \vspace{0.3cm}

\noindent
\textbf{The current $\mathbf{\kappa\theta}$,  $\mathbf{\kappa\theta_0}$} \vspace{0.15cm}

\noindent
$-\frac{4-\sqrt{16+12\Omega_{\theta_0}(w_0+1)(\kappa\delta)^2}}{2\sqrt{3\Omega_{\theta_0}(w_0+1)}}$ \hspace{1cm}
Not a function of $W(\chi)$ \vspace{0.3cm}

\noindent
\textbf{The c parameter} \vspace{0.15cm}

\noindent
$2 (\kappa\delta)^2\ln(\kappa\theta_0)-\kappa\theta_0^2$ \hspace{1cm}
Not a function of $W(\chi)$ \vspace{0.3cm}

\noindent
\textbf{The q parameter} \vspace{0.15cm}

\noindent
$q=-\frac{e^{\frac{c}{(\kappa\delta)^2}}}{(\kappa\delta)^2}$ \vspace{0.3cm}

\noindent
\textbf{$\mathbf{d(\kappa\theta)/da}$} \vspace{0.15cm}

$\frac{-4\kappa\theta}{a((\kappa\theta)^2-(\kappa\delta)^2)}$, \hspace{1cm} $\frac{4\sqrt{-W(\chi)}}{a\kappa\delta(W(\chi)+1)}$
\vspace{0.3cm}

\noindent
\textbf{The beta function} \vspace{0.15cm}

\noindent
$\frac{-4\kappa\theta}{(\kappa\theta)^2-(\kappa\delta)^2}$, \hspace{1cm} $\frac{4 \sqrt{-W(\chi)}} {\kappa\delta(W(\chi)+1)}$
\vspace{0.3cm} 

\noindent
\textbf{$\mathbf{1-\frac{\beta^2(a)}{6}}$} \vspace{0.15cm}

\noindent
$1-\frac{8(\kappa\theta)^2}{3((\kappa\theta)^2-(\kappa\delta)^2)^2}$,\hspace{1cm}  $1+\frac{8W(\chi)}{3\delta^2(W(\chi)+1)^2}$
\vspace{0.3cm}

\noindent
\textbf{The model HI potential} \vspace{0.15cm}

\noindent
$M_m^4((\kappa\theta)^2-(\kappa\delta)^2)^2$, \hspace{1cm} $(M_m\kappa\delta)^4(W(\chi)+1)^2$ \vspace{0.3cm}

\noindent
\textbf{The model potential coefficient $M_m$} \vspace{0.15cm}

\noindent
$\sqrt[4]{\frac{3H_0^2(1-\frac{\beta^2(\kappa\theta_0)}{6}) -\rho_{m_0}}{((\kappa\theta_0)^2-(\kappa\delta)^2)^2}}$, 
\hspace{1.0cm} Not a function of $W(\chi)$  \vspace{0.3cm}      

\noindent
\textbf{The beta potential} \vspace{0.15cm}

\noindent
	$M_{\beta}^4((\kappa\theta)^2-(\kappa\delta)^2)^2(1-\frac{8(\kappa\theta^2}{3((\kappa\theta)^2-(\kappa\theta)^2)^2})$, \hspace{1cm}  $(M_{\beta}\kappa\delta)^4((W(\chi)+1)^2+\frac{8W(\chi)}{3(\kappa\delta)^2})$ \vspace{0.3cm}

\noindent
\textbf{The beta potential coefficient $M_{\beta}$} \vspace{0.15cm}

\noindent
$\sqrt[4]{\frac{3\frac{H_0^2}{\kappa^2}-\frac{\rho_{m_0}}{(1-\frac{8(\kappa\theta^2}{3((\kappa\theta)^2-(\kappa\theta)^2)^2})}}{((\kappa\theta_0)^2-(\kappa\delta)^2)^2}}$  \hspace{1cm}
Not a function of $W(\chi)$ \vspace{0.3cm}

\noindent
\textbf{The Hubble Parameter} \vspace{0.15cm}

$\frac{1}{\sqrt{3}}\sqrt{M_{\beta}^4((\kappa\theta)^2-(\kappa\delta)^2)^2 +\frac{\rho_{m_0}}{a^3(1-\frac{8(\kappa\theta^2}{3((\kappa\theta)^2-(\kappa\theta)^2)^2})}}$,
\hspace{0.25cm}
$\frac{1}{\sqrt{3}}\sqrt{(M_{\beta}\delta)^4(W(\chi)+1)^2+\frac{ \rho_{m_0}}{a^3(1+\frac{8W(\chi)}{3\delta^2(W(\chi)+1)^2})}}$
\vspace{0.3cm}

\noindent
\textbf{The time derivative of $\mathbf{\theta}$, $\mathbf{\dot{\theta}}$} \vspace{0.15cm}

\noindent
$\frac{4\kappa\theta}{\sqrt{3}}\sqrt{M_{\beta}^4+\frac{\rho_{m_0}}{a^3(((\kappa\theta)^2-(\kappa\delta)^2)^2-\frac{8(\kappa\theta)^2}{3})}}$, \hspace{1cm} $\dot{\theta}(a)=\frac{4\sqrt{-W(\chi)}}{\sqrt{3}\kappa\delta}\sqrt{(M_{\beta}\kappa\delta)^4+\frac{\rho_{m_0}}{a^3((W(\chi)+1)^2+\frac{8W(\chi)}{3(\kappa\delta)^2})}}$ \vspace{0.3cm}

\noindent
\textbf{The dark energy density} \vspace{0.15cm}

$\frac{8(\kappa\theta)^2}{3}(M_{\beta}^4 +\frac{\rho_{m_0}}{a^3(((\kappa\theta)^2-(\kappa\delta)^2)^2-\frac{8(\kappa\theta)^2}{3})})+M_{\beta}((\kappa\theta)^2-(\kappa\delta)^2)^2(1-\frac{8(\kappa\theta)^2}{3((\kappa\theta)^2-(\kappa\delta)^2)^2.})$ , \vspace{0.2cm} 

\noindent
$-\frac{8W(\chi)}{3(\kappa\delta)^2}[M_{\beta}\kappa\delta)^4 + \frac{\rho_{m_0}}{a^3((W(\chi)+1)^2 +\frac{8W(\chi)}
{3(\kappa\delta)^2})}] +(M_{\beta}\kappa\delta)^4[(W(\chi)+1)^2 +\frac{8W(\chi)}
{3(\kappa\delta)^2}]$ \vspace{0.3cm}

\noindent
\textbf{The dark energy pressure} \vspace{0.15cm}

\noindent
$\frac{8(\kappa\theta)^2}{3}(M_{\beta}^4 +\frac{\rho_{m_0}}{a^3(((\kappa\theta)^2-(\kappa\delta)^2)^2-\frac{8(\kappa\theta)^2}{3})})-M_{\beta}((\kappa\theta)^2-(\kappa\delta)^2)^2(1-\frac{8(\kappa\theta)^2}{3((\kappa\theta)^2-(\kappa\delta)^2)^2.})$,\vspace{0.2cm}

\noindent
$-(M_{\beta}\kappa\delta)^4(\frac{16W(\chi)}{3(\kappa\delta)^2} + (W(\chi)+1)^2)-\frac{8 \rho_{m_0}W(\chi)}{a^3[3(\kappa\delta)^2(W(\chi)+1)^2+8W(\chi)]}$ \vspace{0.3cm}

\noindent
\textbf{The matter density} \vspace{0.15cm}

\noindent
$\frac{\rho_{m_0}}{a^3}$, \hspace{1cm} Not a function of $W(\chi)$ \vspace{0.3cm}

\noindent
\textbf{The dark energy equation of state} \vspace{0.15cm}

\noindent
$\frac{\frac{8(\kappa\theta)^2}{3}(M_{\beta}^4 +\frac{\rho_{m_0}}{a^3(((\kappa\theta)^2-(\kappa\delta)^2)^2-\frac{8(\kappa\theta)^2}{3})})-M_{\beta}((\kappa\theta)^2-(\kappa\delta)^2)^2(1-\frac{8(\kappa\theta)^2}{3((\kappa\theta)^2-(\kappa\delta)^2)^2.})}{\frac{8(\kappa\theta)^2}{3}(M_{\beta}^4 +\frac{\rho_{m_0}}{a^3(((\kappa\theta)^2-(\kappa\delta)^2)^2-\frac{8(\kappa\theta)^2}{3})})+M_{\beta}((\kappa\theta)^2-(\kappa\delta)^2)^2(1-\frac{8(\kappa\theta)^2}{3((\kappa\theta)^2-(\kappa\delta)^2)^2.})}$  \vspace{0.2cm}

\noindent
$\frac{-(M_{\beta}\kappa\delta)^4(\frac{16W(\chi)}{3(\kappa\delta)^2} + (W(\chi)+1)^2)-\frac{8 \rho_{m_0}W(\chi)}{a^3[3(\kappa\delta)^2(W(\chi)+1)^2+8W(\chi)]}}{-\frac{8W(\chi)}{3(\kappa\delta)^2}[M_{\beta}\kappa\delta)^4 + \frac{\rho_{m_0}}{a^3((W(\chi)+1)^2 +\frac{8W(\chi)}
{3(\kappa\delta)^2})}] +(M_{\beta}\kappa\delta)^4[(W(\chi)+1)^2 +\frac{8W(\chi)}
{3(\kappa\delta)^2}]}$

\end{document}